\documentstyle[11pt,newpasp,twoside,psfig]{article} 
\def\edcomment#1{\iffalse\marginpar{\raggedright\sl#1\/}\else\relax\fi} 
\reversemarginpar 

\begin{document} 
\title{Resolving the 47 Tucanae Distance Problem}

\author{Susan Percival \& Maurizio Salaris} 
\affil{Astrophysics Research Institute, Liverpool John Moores University, UK} 
\author{Francois van Wyk \& David Kilkenny} 
\affil{South African Astronomical Observatory} 

\begin{abstract} 
Using a sample of 43 suitable local subdwarfs with newly acquired $BVI_{C}$ photometry,
we apply our main sequence fitting method to the metal rich Globular Cluster 47 Tucanae.
Fitting in 2 colour planes, we find an apparent distance modulus of 
$(m-M)_{V} = 13.37^{+0.10}_{-0.11}$, leading to a dereddened distance modulus of
$(m-M)_{0} = 13.25^{+0.06}_{-0.07}$.  Consideration of the Red Clump in the cluster 
produces a distance modulus fully consistent with this result.  The implied cluster age
is $11\pm1.4$ Gyr.
\end{abstract}

\section{Introduction} 
Recent empirical distance estimates to 47~Tuc are very discrepant.  MS-fitting by 
several studies (e.g. Carretta et al. 2000) results in an apparent distance modulus
of $(m-M)_{V} > 13.5$, whilst the recent White Dwarf fitting study of Zoccali et al. 
(2001) finds a much shorter distance, $(m-M)_{V}=13.27$.  The results are not consistent
within their quoted errors and the discrepancy in distance moduli implies an uncertainty
in age of $\sim$ 3 Gyr.

Previous MS-fitting studies have suffered from a severe lack of suitable, homogeneous, 
subdwarf (SD) data with which to determine the cluster distance, which in turn leads to 
questions concerning the accuracy of the method.  We have identified a large sample of 
suitable local SDs which enables us to investigate the MS-fitting method thoroughly.
Our multi-waveband data also allows us to fit simultaneously in $V/(B-V)$ and $V/(V-I)$
and because of the different sensitivities of the 2 colours to metallicity and 
reddening, obtaining consistent results is a strong test of the reliability of the 
derived distance.

\section{Data and MS-fitting Method} 
We have acquired $B$, $V$ and $I_{C}$ data for 43 local subdwarfs with $HIPPARCOS$
parallaxes (errors $ < 13 \%$), $M_{V} > 5.5$ and $-1.0 < [Fe/H] < -0.3$. 
Metallicities are
determined from Str\"{o}mgren indices in the literature, and a template (lower) main 
sequence is constructed by applying metallicity dependent colour shifts to each star,
calculated from theoretical isochrones.  
The 47 Tuc main line was derived from the data of Kaluzny et al. (1998), which we
recalibrated using Stetson's `secondary' standards (Stetson 2000).
Full details of sample selection, data collection and the MS-fitting 
method can be found in Percival et al. 2002.

\section{Results}
Assuming $[Fe/H] = -0.7\pm0.1$ and $E(B-V)=0.04\pm0.02$ for the cluster, our best-fit
distance modulus in both colour planes is $(m-M)_{V}=13.37^{+0.10}_{-0.11}$ (figure 1),
leading to a dereddened distance modulus of $(m-M)_{0}=13.25^{+0.06}_{-0.07}$.
From the turn off luminosity, the implied cluster age is $11\pm1.4$ Gyr.

\begin{figure} 
\psfig{figure=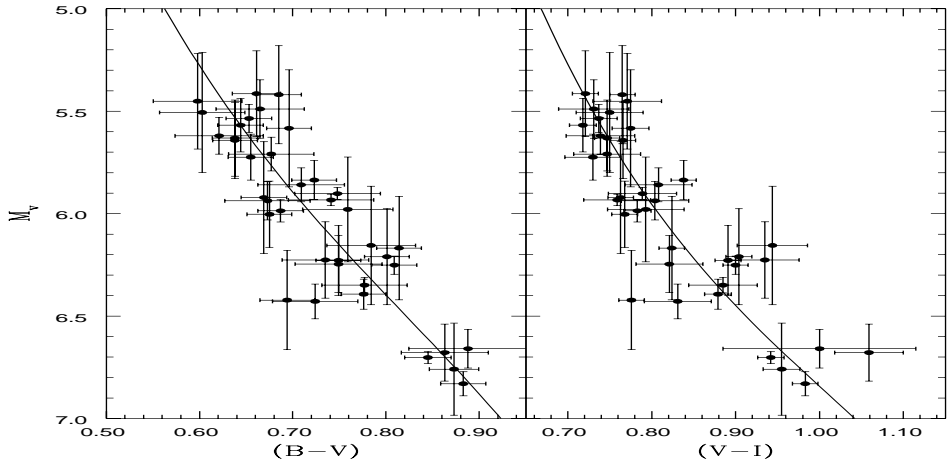,height=6.5cm,width=13.3cm,clip=}
\caption{Subdwarfs `shifted' to $[Fe/H]=-0.7$ and the recalibrated 47 Tuc main 
lines, using $(m-M)_{V}=13.37$.}
\end{figure}

Independent support for our MS-fitting distance comes from consideration of the Red 
Clump (RC) as a distance indicator.  The red horizontal branch of 47~Tuc is the 
counterpart of the RC in the solar neighbourhood, whose $I$-band absolute magnitude is 
precisely determined by {\it HIPPARCOS} parallaxes.  Applying the appropriate 
evolutionary correction from Girardi \& Salaris~(2001) to the cluster RC, we derive a
dereddened distance modulus of $(m-M)_{0,RC}=13.31\pm0.05$, which agrees well with our
MS-fitting result.

\end{document}